# ACCELERATION RATES AND INJECTION EFFICIENCIES IN OBLIQUE SHOCKS

Donald C. Ellison

Department of Physics, North Carolina State University,

Box 8202, Raleigh NC 27695, U.S.A.

don_ellison@ncsu.edu

Matthew G. Baring[1] and Frank C. Jones

Laboratory for High Energy Astrophysics, Code 665,

NASA Goddard Space Flight Center, Greenbelt, MD 20771, U.S.A.

baring@lheavx.gsfc.nasa.gov, jones@lheavx.gsfc.nasa.gov

## ABSTRACT

The rate at which particles are accelerated by the first-order Fermi mechanism in shocks depends on the angle, $\Theta_{\mathrm{Bn1}}$, that the upstream magnetic field makes with the shock normal. The greater the obliquity the greater the rate, and in quasi-perpendicular shocks (i.e., $\Theta_{\mathrm{Bn1}} \to 90°$) rates can be hundreds of times higher than those seen in parallel shocks. In many circumstances pertaining to evolving shocks (e.g., supernova blast waves and interplanetary traveling shocks) or where acceleration competes with losses (e.g., through synchrotron cooling), high acceleration rates imply high maximum particle energies and obliquity effects may have important astrophysical consequences. However, as is demonstrated in this paper, the efficiency for injecting thermal particles into the acceleration mechanism also depends strongly on obliquity and, in general, varies inversely with $\Theta_{\mathrm{Bn1}}$; shocks which accelerate particles most rapidly are *least* capable of injecting thermal particles into the acceleration process. In addition, the degree of turbulence and the resulting cross-field diffusion strongly influences both injection efficiency and acceleration rates. The test particle Monte Carlo simulation of shock acceleration used here assumes large-angle scattering, computes particle orbits exactly in shocked, laminar, non-relativistic flows (contrasting our previous simulations of oblique shocks, e.g., Baring et al. 1994, which assumed magnetic moment conservation), and calculates the injection efficiency as a function of obliquity, Mach number, and degree of turbulence. We find that turbulence must be quite strong for high Mach number, highly oblique shocks to inject significant numbers of thermal particles and that only modest gains in acceleration rates can be expected for strong oblique shocks over parallel ones if the only source of seed particles is the thermal background.

*Subject headings:* Cosmic rays: general — particle acceleration — shock waves — diffusion

---

## 1. INTRODUCTION

It has been shown in analytic test-particle calculations (i.e., Jokipii 1987; Ostrowski 1988b) that the Fermi acceleration rate in oblique shocks can be substantially greater than that in quasi-parallel ones; here quasi-parallel shocks are defined as those where the angle $\Theta_{Bn1}$ between the shock normal and the ambient upstream magnetic field is approximately zero. Since in some circumstances, such as supernova remnant (SNR) blast waves, the maximum energy obtained by the accelerated particles scales as the acceleration rate, the rate of acceleration can have important astrophysical significance. It has been argued that the large increase in maximum energy obtainable in nearly perpendicular shocks may solve the long standing problem of how (or if) cosmic rays up to the so-called 'knee' at $\sim 10^{15}$ eV can be produced in SNRs (Jokipii 1987) and explain the origin of the highest energy cosmic rays in a galactic wind termination shock (Jokipii and Morfill 1985).

In this paper, we reexamine the variation of acceleration rate with shock obliquity using a test-particle Monte Carlo simulation which allows a determination of *both* the acceleration rate and the injection efficiency from a thermal background as a function of obliquity, Mach number, and degree of turbulence (i.e., cross-field diffusion). While we confirm previous results of rapid acceleration when $\Theta_{Bn1} \to 90°$, we find that there is an inverse correlation between acceleration rate and injection efficiency from the thermal background; shocks which accelerate particles rapidly inject few thermal particles. Both the acceleration rate and the injection efficiency depend strongly on the amount of scattering and the Mach number. In general, the higher the Mach number and weaker the scattering, the fewer particles injected. For the high Mach numbers expected in young SNR blast waves, the oblique portions of the shocks which accelerate particles rapidly do not pick up significant numbers of particles from the ambient thermal interstellar medium unless there is strong scattering. Of course, rapid acceleration may occur if superthermal seed particles are present or if the shock geometry is such that significant numbers of thermal particles accelerated at quasi-parallel portions of the shock are subsequently injected and accelerated at quasi-perpendicular regions.

The simulation used here is an adaptation of our Monte Carlo technique which has been described in numerous papers including Baring, Ellison and Jones (1993, 1995). The technique uses phenomenological, large-angle scattering such that a particle's direction is isotropized in a single scattering event with a mean free path for collisions proportional to a particle's momentum. We believe such a scattering law is a good approximation to that determined by hybrid plasma simulations (e.g., Max et al. 1988; Zachary 1987; Giacalone et al. 1993) and also to that inferred from spacecraft measurements at shocks in the heliosphere (e.g., Ellison, Möbius, and Paschmann 1990; Giacalone, Burgess and Schwartz 1992). In any case, our parameterization of wave-particle interactions is a simplification that renders the Monte Carlo technique extremely amenable to modeling the acceleration properties of oblique shocks, and indeed it generates information on particle distributions and acceleration efficiencies much faster than cumbersome plasma simulations where trajectories are calculated from electric and magnetic forces. The potential necessity for using 3-D plasma codes in modeling shocks (e.g., Jokipii, Kóta and Giacalone 1993), as discussed in Section 3 below, further enhances the usefulness of the Monte Carlo approach, where diffusion is intrinsically three-dimensional. While former applications of our oblique shock techniques (e.g. Baring, Ellison and Jones 1993, 1994; Baring et al. 1995) invoked a guiding center approximation, where the details of particle gyrations about their gyrocenter were ignored and conservation of magnetic moment was imposed, here we calculate orbits exactly as in the diverse works of Decker and Vlahos (1985), Decker (1988), Ostrowski (1988a, 1991), Takahara and Terasawa (1991) and Begelman and Kirk (1990), and make no assumption relating to the magnetic moment. The code, as presented here, is only intended for subluminal cases and any reference to quasi-perpendicular shocks (unless clearly stated otherwise) will imply shocks with large $\Theta_{Bn}$, but which are still subluminal.

We also note that although our oblique Monte Carlo simulation includes injection from thermal energies (which is omitted in the earlier gyrohelix treatments of Decker and Vlahos 1985; Decker 1988; Ostrowski 1988a, 1991; Begelman and Kirk 1990; and Takahara and Terasawa 1991) and a parameterization of cross-field diffusion, it is still a test-particle calculation in that the shock transition is assumed to be discontinuous on all length scales. While not applicable to high Mach number shocks where significant particle acceleration can be expected, we believe such a test-particle picture is appropriate for modeling oblique interplanetary traveling shocks, which have low Mach numbers, small compression ratios and therefore relatively little pressure in the accelerated population. Indeed, the application of our simulation to such scenarios (Baring et al. 1995) has indicated the importance of field turbulence in achieving the efficient injection required to match simulation output with observational data for the non-thermal particle distributions. In contrast, it has been suggested for high Mach number quasi-parallel shocks (e.g., Eichler 1984, 1985; Ellison and Eichler 1984) that Fermi acceleration is self-regulating and can place a large fraction of the total energy flux into accelerated particles even if the injection rate out of the thermal pool is small (e.g., Ellison 1985). This is particularly relevant to the quasi-parallel regions of the earth's bow shock, where good agreement between simulation spectra and observations is only obtained (Ellison, Möbius, and Paschmann 1990) when including the pressure of the accelerated population in the determination of the hydrodynamic structure of the shock; the shock transition is broadened by the backpressure of accelerated particles. We are currently working to generalize our simulation to include this self-regulation in oblique shock geometries.

Finally, note that we have previously presented preliminary results on oblique shock injection using a guiding center approach (Baring, Ellison, and Jones 1995), and work is in progress to compare the present exact orbit calculations to our guiding center version of the Monte Carlo technique.

## 2. ANALYTIC ESTIMATION OF THE ACCELERATION RATE

Before presenting the determination of acceleration rates using the Monte Carlo simulation, it is instructive and appropriate to review Jokipii's (1987, see also Ostrowski 1988b) analytic derivation of the acceleration time and its adaptation to our simulation technique. Jokipii showed how the acceleration rate can be calculated when the effects of particle drifts are included in the diffusive theory of Fermi acceleration (see Drury 1983; Blandford and Eichler 1987; and Jones and Ellison 1991 for reviews of the basic Fermi mechanism). He demonstrated that as $\Theta_{\mathrm{Bn1}}$ approaches $90°$, the acceleration becomes much more rapid than in quasi-parallel shocks (i.e., when $\Theta_{\mathrm{Bn1}} \simeq 0°$). Further he indicated that most of the energy gain in Fermi acceleration in quasi-perpendicular shocks comes from particles drifting in the $\mathbf{u} \times \mathbf{B}$ electric field, where $\mathbf{u}$ is the local flow velocity and $\mathbf{B}$ is the local magnetic field. This drifting acceleration, which is often called the 'shock-drift' mechanism (e.g., Decker 1988), is a fundamental part of diffusive Fermi acceleration and is contained in the standard diffusive formulation of the theory (e.g., Blandford and Ostriker 1978; Drury 1983).

We note that by transforming to the de Hoffmann-Teller frame of reference (de Hoffmann and Teller, 1950), where the shock is stationary and $\mathbf{u} \times \mathbf{B} = \mathbf{0}$ everywhere, the electric field is identically zero everywhere. Hence it follows that the so-called shock drift mechanism is inseparable from, and intrinsically part of, the Fermi process (e.g., Drury 1983; Jones and Ellison, 1991); it is automatically included in our Monte Carlo technique (Baring, Ellison and Jones, 1993), which treats particle transport in the de Hoffmann-Teller (HT) frame using a guiding-center approach to particle transport.

We summarize Jokipii's derivation by first writing down the standard form for the acceleration time, $\tau_{\mathrm{a}}$, to a given momentum $p$, which is derivable directly (e.g., Forman and Morfill 1979; Drury 1983) from the diffusion equation:

$$\tau_{\mathrm{a}}(p) = \frac{3}{u_{1x} - u_{2x}} \int_{p_{\mathrm{i}}}^{p} \left( \frac{\kappa_1}{u_{1x}} + \frac{\kappa_2}{u_{2x}} \right) \frac{dp'}{p'} \quad , \tag{1}$$

where $p_{\mathrm{i}}$ is the momentum of the injected particles, $u_{1x}$ ($u_{2x}$) is the upstream (downstream) component of flow speed normal to the shock in its rest frame, and $\kappa_1$ and $\kappa_2$ are the upstream and downstream spatial diffusion coefficients in the direction normal to the shock. The subscript 1 (2) will always imply quantities determined upstream (downstream) from the shock, and the negative $x$-direction will denote the shock normal. The form in Eq. (1) depends only on the assumption of virtual isotropy of the cosmic rays in the shock frame, which arises only for particle speeds far in excess of the flow speeds $u_{1x}$ and $u_{2x}$. It therefore follows that this diffusive formula cannot be applied to relativistic shocks where the particle distribution is usually quite anisotropic at most energies of interest; deviations from it in the relativistic domain are clearly depicted by Ellison, Jones and Reynolds (1990). Following Jokipii (1987), hereafter the so-called normal incidence frame (NIF), where the upstream flow is along the shock normal, is chosen as the reference perspective.

The diffusion coefficients $\kappa_1$ and $\kappa_2$ can then be written down following appropriate rotation of the elements of the diffusion tensor from coordinates oriented with the magnetic field lines:

$$\begin{aligned} \kappa_1 &= \kappa_{1\|} \cos^2 \Theta_{Bn1} + \kappa_{1\perp} \sin^2 \Theta_{Bn1} \quad , \\ \kappa_2 &= \kappa_{2\|} \cos^2 \Theta_{Bn2} + \kappa_{2\perp} \sin^2 \Theta_{Bn2} \quad . \end{aligned} \qquad (2)$$

Here $\kappa_\|$ and $\kappa_\perp$ are the diffusion coefficients respectively parallel and perpendicular to the average magnetic field (calculated both upstream and downstream from the shock); non-zero values of $\kappa_\perp$ amount to the presence of so-called *cross-field diffusion*. Note that $\Theta_{Bn2}$ is the angle between the shock normal and the downstream magnetic field.

Jokipii (1987) related $\kappa_\perp$ to $\kappa_\|$ using the standard kinetic theory result (e.g., Axford 1965; Forman, Jokipii and Owens 1974):

$$\kappa_\perp = \frac{\kappa_\|}{1 + (\lambda/r_g)^2} \quad , \qquad (3)$$

where $\lambda$ is the scattering mean free path, $r_g = pc/(qB)$ is the gyroradius, $c$ is the speed of light, $q = eZ$ is the charge of an ion with charge state $Z$, and the mean magnetic field $B$ will change from $B_1$ when upstream to $B_2$ downstream. This identity is tantamount to shifting a particle by approximately one gyroradius in *any direction* during scatterings with the turbulent magnetic field. Jokipii (1987) observed that there is no general consensus as to the form of the relationship between $\kappa_\perp$ and $\kappa_\|$, particularly if the effects of field-line meandering are to be included in the diffusion. Therefore, the choice of the ratio is somewhat subjective, though the requirement that a particle move of the order of a gyroradius in a scattering must always be satisfied. The form in Eq. (3) elicits the property that the Bohm diffusion (strong scattering) limit, defined by $\lambda \sim r_g$, yields $\kappa_\perp \sim \kappa_\|$. Further, $\kappa_\perp \ll \kappa_\|$ in the limit of $\lambda/r_g \to \infty$.

In this paper, the adaptation of our Monte Carlo simulation that is used to generate test-particle distributions, acceleration rates, and efficiencies is a gyro-orbit computation, where we determine the exact gyrohelices describing particle motion between scatterings in uniform fields numerically. Such "gyrohelix" calculations have been performed, for example, by Decker (1988, and see references therein) and Takahara and Terasawa (1991) for non-relativistic shocks, and by Begelman and Kirk (1990) for relativistic, superluminal shocks. The choice of the scattering "operator" is, of course, subjective, and we opt for a scenario of large-angle scattering, where the momentum of a particle is isotropized upon scatterings. This contrasts pitch-angle diffusion (e.g. see Decker and Vlahos 1985; Ostrowski 1988a, 1991), and is chosen to mimic the effect of large amplitude field turbulence on the particles. This implementation of such large-angle scattering in our simulation automatically implies the applicability of the kinetic theory result in Eq. (3) to our simulation.

To complete our formalism, we set (following Jokipii, 1987) $\lambda$ to be some constant number of gyroradii, i.e., $\lambda = \eta r_g$, and, therefore, $\kappa_\| = \eta r_g v/3$, where $v$ is the particle speed measured in the local plasma frame. Since we take $\eta$ to be the same upstream and downstream from the shock (Jokipii made this assumption also), $\lambda$ is implicitly assumed to be inversely proportional to $B$. These subjective choices can be altered whenever desired, and indeed the Monte Carlo technique can accommodate any form for $\lambda$ as a function of $p$.

Using the above definitions in Eqs. (1) and (3), and remembering that the component of $B$ normal to the shock is conserved across it (i.e., $B_1 \cos\Theta_{Bn1} = B_2 \cos\Theta_{Bn2}$), the acceleration time as a function of shock obliquity and Mach number, $\tau_a(\Theta_{Bn1}, M_1)$, can be written as

$$\tau_a(\Theta_{Bn1}, M_1) = \frac{\eta c}{qB_1} \frac{E - E_i}{u_{1x} - u_{2x}} \left\{ \frac{1}{u_{1x}} \left[ \cos^2\Theta_{Bn1} + \frac{\sin^2\Theta_{Bn1}}{1+\eta^2} \right] \right.$$
$$\left. + \frac{1}{u_{2x}} \frac{\cos\Theta_{Bn2}}{\cos\Theta_{Bn1}} \left[ \cos^2\Theta_{Bn2} + \frac{\sin^2\Theta_{Bn2}}{1+\eta^2} \right] \right\} \quad (4)$$

where the kinetic energies $E$ and $E_i$ correspond to $p$ and $p_i$. The upstream Mach number, $M_1$, is determined from $M_1^{-1} = M_{s1}^{-1} + M_{A1}^{-1}$, where $M_{s1} = [\rho_1 u_{1x}^2/(\gamma P_1)]^{1/2}$ is the upstream sonic Mach number and $M_{A1} = v_{A1}/u_1$ is the upstream Alfvén Mach number. Here $\rho_1$ is the upstream density, $\gamma = 5/3$ is the ratio of specific heats, $P_1$ is the upstream pressure, and $v_{A1} = B_1/\sqrt{4\pi\rho_1}$ is the Alfvén speed in the upstream region. The Mach number dependence (and hence implicitly the dependence on magnetic field) in Eq. (4) enters through the compression ratio $r = u_{1x}/u_{2x}$ and through $\Theta_{Bn2}$, which depends on the plasma $\beta = nkT/[B^2/(8\pi)]$ ($n$ is the number density and $k$ is Boltzmann's constant) and must be determined from the standard Rankine-Hugoniot relations (e.g., see Decker 1988; Jones and Ellison 1991). Equation (4) is relativistically correct and note that as long as $E \gg E_i$, $\tau_a$ is simply proportional to $E$: this functional form depends only on the fact that $\lambda$ was chosen to be proportional to $p$. Note also that Eq. (4) bears close resemblance to Eq. (9) of Ostrowski (1988b), who used a different parameterization of the spatial diffusion coefficients.

As an illustration, the expression for $\tau_a(\Theta_{Bn1}, M_1)$ in Eq. (4) is depicted (smooth curves) in Fig. 1 for $\Theta_{Bn1} = 60°$ and for $\eta = \lambda/r_g = 3$ and 100, indicating that the acceleration time to a particular energy increases substantially with increasing $\eta$. For nearly perpendicular shocks this behavior reverses and $\tau_a$ decreases as $\eta$ increases. Therefore, $\Theta_{Bn1}$ has a profound influence on the acceleration time and this fact is explicitly illustrated in Figs. 2 and 3. We note that Eq. (4) is correct for all values of $B$ even though we have chosen, in most of the calculations presented below, to use a small value of $B$ so that shocks with different $\Theta_{Bn1}$ will have the same compression ratio (in Fig. 1, $r \simeq 4$ for both cases). The histograms in Fig. 1 are obtained from the Monte Carlo simulation and are discussed below. Note that the $\eta = 100$ case corresponds to little cross-field diffusion, while $\eta = 3$ corresponds to quite strong scattering (as mentioned above, $\eta = 1$ is the so-called Bohm limit). When $\Theta_{Bn1} = 0°$ and for $\lambda = \eta r_g$, $\tau_a$ scales as $\eta$ and inversely as $B$ but depends only weakly on Mach number unless $M_1$ drops below about 3, and the compression ratio drops significantly below the strong shock value of $r \simeq 4$.

The principal quantity of interest in Jokipii (1987) was the ratio, $R_a$, of the acceleration rate for a shock of a given obliquity to that of a parallel shock ($\Theta_{Bn1} = 0°$); this may be obtained directly from the inverse of the acceleration time,

$$R_{\rm a} \;=\; \frac{\tau_{\rm a}(0,\,M_1)}{\tau_{\rm a}(\Theta_{\rm Bn1},\,M_1)} \quad , \tag{5}$$

and suitably defines the variation of acceleration rate with shock obliquity, and also with $\eta = \lambda/r_{\rm g}$. This result for $R_{\rm a}$ differs somewhat from that obtained in Eq. (8) of Jokipii (1987): we calculate $R_{\rm a}$ with $B_1 = B_2$ in the numerator of Eq. (5), which is appropriate for plane-parallel shocks, whereas Jokipii (1987) used the $B_1/B_2$ ratio determined for $\Theta_{\rm Bn1} \neq 0$ in both numerator and denominator. Our result therefore changes the numerator in Jokipii's Eq. (8) simply to $1 + r$, and makes our $R_{\rm a}$ somewhat larger than his.

In Fig. 2 we depict the ratio $R_{\rm a}$, as determined from Eq. (5), as a function of $\Theta_{\rm Bn1}$ for different $\eta$ (solid curves). These curves clearly indicate that the acceleration time *ratio* is a strong function of shock obliquity and $\eta$. Also shown in Fig. 2 as dashed lines are the results for $R_{\rm a}$ using the alternative prescription of Jokipii (1987), for $\eta = 1$ and 100: these fall below our results for the reasons just mentioned (note the large difference when $\eta = 1$). It is clear that the acceleration rate (compared to $\Theta_{\rm Bn1} = 0°$) is largest when cross-field diffusion is weak (i.e., when $\eta$ is large) and when $\Theta_{\rm Bn1}$ is near $90°$. However, it is the maximum energy obtainable in an actual shock that is astrophysically important (e.g. see Lagage and Cesarsky 1983; Jokipii and Morfill 1987), not $R_{\rm a}$! If the system is evolving, or if energetic particles lose energy through some competing process such as synchrotron emission, the smaller $\tau_{\rm a}$ becomes, the higher the energy obtainable becomes. Therefore, $\eta$ cannot be made arbitrarily large to increase $R_{\rm a}$ since the acceleration time scales as $\eta$ (note this scaling in Fig. 1), and real systems will require substantial turbulence to accelerate particles to high energies.

To clarify this, we plot in Fig. 3 the acceleration time to a given energy (in this case $10^3$ keV) versus $\Theta_{\rm Bn1}$ for different $\eta$. Results are normalized by the upstream magnetic field strength. Two aspects of the dependence of the acceleration time become immediately apparent. Firstly, the acceleration time is quite insensitive to Mach number as long as $M_1 \gtrsim 3$ (this is true regardless of $\eta$ although only the $\eta = 100$ case is shown in Fig. 3). For low magnetic fields, the acceleration time depends on Mach number mainly through the $u_{1x} - u_{2x}$ term in Eq. (4), i.e. via the compression ratio. Secondly, over most of the range in $\Theta_{\rm Bn1}$ the value of $\eta$ determines the acceleration time and only when $\Theta_{\rm Bn1} \gtrsim 75°$ and $\eta \gtrsim 10$ does the dependence on obliquity begin to dominate. It is clear that for the large majority of phase space, the acceleration will be most rapid when scattering is strong (i.e. $\eta \lesssim 10$). The representation of acceleration times in Fig. 3 is of greater relevance to (and provides more insight for) astrophysical problems than the acceleration rate ratios in Fig. 2 (and of Jokipii 1987), and was adopted by Ostrowski (1988b), who focused on cases where $\eta \gg 1$. Note that the acceleration time in Eq. (4) should clearly be proportional to $\eta = \lambda/r_g$, since while the gyrofrequency is the fundamental timescale for particle motion in a field, the scattering time $\eta c/(qB)$ principally forms the scaling for the time particles take to complete a given number of shock crossings, and therefore attain a certain energy. This dependence defines the sensitivity of $\tau_{\rm a}$ in Fig. 3 to $\eta$ in the regime of quasi-parallel shocks. If the acceleration time is scaled by $\eta$ to form a time for acceleration to a given energy *per scattering*, then the curves in Fig. 2 clearly indicate that the time per scattering is a decreasing function of $\eta$.

The analytic results of Eqs. (4) and (5) contain no information relating to the thermal particles and give no indication of how efficiently the high energy non-thermal population is produced; below it is shown that both turbulence and Mach number have strong influences on the injection efficiency from the thermal background. While Jokipii (1987) did not specifically consider injection aspects of the acceleration problem, he remarked that low-energy injection would be more probable in quasi-parallel shocks. Jokipii emphasized that acceleration time results like those in Eqs. (4) and (5) are valid only when the diffusion length scale $\kappa_1/u_{1x}$ exceeds the particle gyroradius (see also Jones and Ellison, 1991). In quasi-perpendicular shocks, the form in Eq. (1) can then be used to deduce the criterion

$$\eta \lesssim \frac{v}{u_{1x}} \qquad (6)$$

for validity of Eq. (4). The regime of quasi-perpendicularity is denoted by $\kappa_{1\perp}\sin^2\Theta_{\mathrm{Bn1}} \gtrsim \kappa_{1\|}\cos^2\Theta_{\mathrm{Bn1}}$, i.e. $\eta \lesssim \tan\Theta_{\mathrm{Bn1}}$ when $\eta \gg 1$, and indeed Eq. (6) can be easily generalized using Eq. (2) to arbitrary obliquities. Clearly Eq. (6) becomes difficult to satisfy for quasi-thermal particles ($v \sim u_{1x}$), except in the Bohm diffusion limit.

Consider now the diffusion "velocity" $\kappa_{1\perp}\nu_s/u_{1x}$ perpendicular to the field, where $\nu_s = \Omega/\eta$ is the scattering frequency and $\Omega$ is the gyrofrequency. Since $\kappa_{1\perp} = v^2/[3u_{1x}(1+\eta^2)]$, using Eq. (3), where $v$ is the speed of the particle, it is easily shown that the ratio of perpendicular diffusion velocity to convection speed $u_{1x}$ varies roughly as the inverse of the square of the parameter $\eta u_{1x}/v$ when $\eta \gg 1$. Jokipii (1987) dubbed this parameter the *streaming anisotropy*, so that the criterion in Eq. (6) just corresponds to quasi-isotropic streaming. We argue here that this streaming condition is essential in quasi-perpendicular shocks for efficient injection of particles into the acceleration process from the thermal population. The injection of thermal particles can only be efficient when these particles can return to the shock with significant probability after crossing to the downstream side of it for the first time. This probability of return is substantial only when the streaming anisotropy is small, thereby yielding Eq. (6) as the requirement for efficient injection at quasi-perpendicular shocks, or more specifically $\eta \sim 1$ since $v \sim u_{1x}$ at thermal energies. This requirement is borne out in the results of the simulations that are presented below.

## 3. THE MONTE CARLO TECHNIQUE

In previous papers (Baring, Ellison and Jones 1993, 1994, 1995) we have described our Monte Carlo technique for calculating particle acceleration in oblique shocks, which is a kinematic model that closely follows Bell's (1978) approach to diffusive acceleration. The technique is to inject thermal particles in the region upstream of an oblique shock, mimicking for example solar wind ions, and allow them to convect in the flow and scatter, crossing the shock a few or many times before they eventually leave the system either far downstream or when they achieve an energy in excess of some specified maximum; typical particle trajectories are similar to those illustrated in Baring, Ellison and Jones (1994). Scattering is included phenomenologically by exponentially distributing the scattering time about the average collision time, $t_c = \eta/\Omega$, where $\Omega = ZeB/(Am_p c)$ is the ion gyrofrequency ($B$ is the field strength on either side of the shock, $A$ is the mass number, and $m_p$ is the proton mass). The scattering is elastic and isotropizes the particles in the local fluid frame in a single scattering event, thereby mimicking the effect of large amplitude field turbulence on particle motions. Such turbulence is present in both plasma simulations (e.g. Quest 1988; Burgess 1989; Winske et al. 1990) and observations of shocks in the heliosphere (e.g., Hoppe et al. 1981), motivating our choice; we remark that our technique is adaptable (e.g. see Ellison, Jones and Reynolds 1990) to treat the pitch angle diffusion that is assumed by Decker and Vlahos (1985) and Ostrowski (1988a, 1991). The mean free path for collisions is chosen to be proportional to a particle's momentum, a scattering law that is a reasonable approximation to data from hybrid plasma simulations (Giacalone, Burgess and Schwartz 1992; Giacalone et al. 1993) and also to the scattering inferred from measurements at the earth's bow shock (Ellison, Möbius, and Paschmann 1990).

All of these properties are included in the present adaptation of the Monte Carlo technique. However, former applications of this method (e.g. Baring, Ellison and Jones 1993, 1994; Baring et al. 1995) invoked a guiding center approximation, where the details of particle gyrations about their gyrocenter were ignored, and the adiabatic approximation was invoked in shock crossings. Here, we omit this simplification, calculate orbits exactly, and allow the particles to gyrate between scatterings and during their interaction with the shock discontinuity. No assumption about conservation of magnetic moment is made. Gyrohelix approaches are common in the literature (e.g. see Decker, 1988 for a review), and we note that ours bears closest resemblance to the treatment of Takahara and Terasawa (1991), though they do not consider particle injection or regimes of strong scattering ($\eta \sim 1$). We remark that work is in progress to compare the present orbit calculations with the results on oblique shock injection using the guiding center approach that are described in Baring, Ellison and Jones (1995). Further, we note that in regimes of weak scattering, such as for the case of Fig. 5, any imposition of the adiabatic approximation is generally quite accurate for quasi-perpendicular shocks (e.g. Drury, 1983), mimicking the principal characteristics of gyrohelix calculations (e.g. Terasawa, 1979).

The test-particle Monte Carlo simulation used here moves particles in the HT frame by calculating the orbit assuming a uniform background magnetic field with strength $B_1$ upstream and $B_2$ downstream. The HT frame is determined by moving parallel to the shock front with a speed equal to the speed the intersection point of the magnetic field with the shock moves (i.e., $v_{\mathrm{HT}} = u_1 \tan\Theta_{\mathrm{Bn1}}$). In this frame, $\mathbf{u}$ and $\mathbf{B}$ are parallel and there is no $\mathbf{u} \times \mathbf{B}$ electric field. At the shock the field kinks such that the angle between the magnetic field and the shock normal goes from $\Theta_{\mathrm{Bn1}}$ upstream to $\Theta_{\mathrm{Bn2}}$ downstream, but in the HT frame $\mathbf{u}$ and $\mathbf{B}$ remain parallel both upstream and downstream. For a given set of upstream parameters (i.e., $u_1$, $T_1$, $n_1$, $B_1$, and $\Theta_{\mathrm{Bn1}}$), the standard Rankine–Hugoniot relations are used to determine the downstream values (e.g., Decker 1988; Jones and Ellison, 1991). The gyrohelix equations are stated in numerous papers (e.g., see Terasawa 1979; Begelman and Kirk 1990; Baring and Riitano 1995), as are the transformations associated with conservation of particle momentum across the shock. A comprehensive treatment of particle transmission and reflection distributions and associated energy changes at oblique shocks can be found in Baring and Riitano (1995).

Our procedures for determining acceleration rates for parallel shocks are described in Ellison, Jones, and Reynolds (1990) and these have been updated here to include oblique shocks. In the simulation, the acceleration time is the total time a particle spends in the system starting when it first crosses the shock until it reaches an energy $E$, including time spent both upstream and downstream from the shock. The only subtlety is how the time is calculated when the particles are far downstream of the shock. To render the simulation finite in time, while maintaining good population statistics, it is necessary to introduce a downstream "boundary" to the simulation region, beyond which the spatial diffusion properties are treated using appropriate statistical probabilities. When particles cross to the downstream side of this boundary, which is more than a scattering length downstream of the shock, their probability of return due to convection and spatial diffusion is determined via the well-known formula (e.g. see Jones and Ellison 1991) $(v - u_{2x})^2/(v + u_{2x})^2$, where $v$ is the particle speed (in the plasma frame) at the point of crossing the boundary. The decision of return (or otherwise) is made via a random number generator. If the particle does return to the boundary, its pitch angle is appropriately distributed, taking into account flux-weighting, and since the scattering length is exponentially-distributed about the mean free path, the particle can be simply relocated at the downstream boundary without introduction of a spatial bias. Further, the contribution to the particle's acceleration time incurred downstream of this boundary is determined by the "retrotime" technique described in Ellison, Jones, and Reynolds (1990): after a particle returns upstream of the boundary, it is followed backward in time. Essentially, the velocity of the particle is reversed, and so is the flow, so that a *probable* history is tracked and an estimate for the average time of return is obtained. Returning particles from this downstream boundary in such a fashion mimics a downstream region of infinite extent.

It must be noted that our parameterization of wave-particle interactions could be quite different from what actually occurs in oblique shocks where magnetic field overshoots and other microphysical effects are known to strongly influence low energy particles (e.g., see Quest 1988; Scholer, Fujimoto and Kucharek 1993 for quasi-parallel shocks, and Winske and Quest 1988 for quasi-perpendicular shocks). While these processes are best studied using plasma simulations, it has recently been shown both analytically by Jokipii, Kóta and Giacalone (1993) and numerically by Giacalone and Jokipii (1994) that plasma simulations of highly oblique shocks must be fully three-dimensional in order to self-consistently include cross-field diffusion. To our knowledge, no such 3-D simulations that show injection rates or significant particle acceleration have been reported. Indeed, the comparison of 1-D and 2-D hybrid simulations of quasi-parallel shocks (Kucharek, Fujimoto and Scholer 1993) indicate that the dimensionality of the simulation is quite influential on the results obtained. Since an accurate determination of injection from plasma techniques will require simulations that are run long enough to generate well-developed wave turbulence in the foreshock region, it may be a long time before these three-dimensional simulations are implemented. While our method parameterizes scattering, it incorporates the three-dimensional nature of wave-particle collisions and goes beyond any current analytic attempt to treat injection. Until the computational requirements of plasma simulations can be met, our technique serves as a valuable intermediate step between analytic calculations and 3-D plasma simulations.

In principle, our Monte Carlo method is identical to the analytic estimates discussed in Section 2, except that the simulation does not rely on the diffusion approximation, as does the derivation of Eq. (1), making it possible for us to treat both thermal and superthermal particles. While this will have virtually no effect on the acceleration times to energies well above thermal, it does allow us to estimate the efficiency with which thermal particles are injected into the acceleration process as a function of $\Theta_{\rm Bn1}$, $\eta$, and Mach number.

## 4. ACCELERATION TIMES AND EFFICIENCIES FROM SIMULATIONS

The Monte Carlo simulation used here effectively solves the test-particle Boltzmann equation describing Fermi shock acceleration. At energies large enough so that the diffusion approximation is valid [i.e., $v/u_{1\rm HT} \gg 1$, where $u_{1\rm HT} = u_{1x}/\cos\Theta_{\rm Bn1}$ is the upstream flow speed in the HT frame], it yields the standard test-particle Fermi power law. However, since no approximation to large speeds is needed for the Monte Carlo technique, we also calculate the particle distribution function at thermal energies as well and, therefore, obtain a parameterization of injection at energies where the diffusion approximation is not valid. In Fig. 4 we show an example of our results for parameters typical of shocks in the interstellar medium, i.e., $u_{1x} = 500$ km s$^{-1}$, $T_1 = 10^6$ K, $\Theta_{\rm Bn} = 30°$, $\eta = 3$, $B_1 = 2 \times 10^{-5}$ G, and $n = 1$ cm$^{-3}$. These values produce a shock with $M_{\rm s1} = 4.3$, $M_{\rm A1} = 11.5$, $\beta = 8.7$, $r = 3.4$, and $\Theta_{\rm Bn2} = 63°$. The omni-directional number flux, $dJ/dE$, [in particles/(cm$^{-2}$-s-sr-keV)] obtained from the simulation is plotted in Fig. 4 as a solid histogram, and at energies such that $v \gg u_{1\rm HT}$ ($\sim 300$ keV for this particular example) approximates, within statistical uncertainty, the power-law predicted by the standard Fermi theory (e.g. Jones and Ellison 1991), i.e., $dJ/dE \propto E^{-[(r+2)/(r-1)]/2}$. The accuracy of the reproduction of this form by the Monte Carlo simulation is illustrated in Fig. 4 by also plotting $E^{[(r+2)/(r-1)]/2}\,dJ/dE$ (dashed line), which will be horizontal if the predicted slope is obtained. A least squares fit to the Monte Carlo power law above 300 KeV yields, $\sigma = 1.13 \pm 0.02$ compared to a predicted value of $\sigma = 1.12$. This example is not intended to model any particular heliospheric shock but is used to illustrate that our technique reproduces the test-particle Fermi prediction at large energies for any combination of parameters (including large magnetic fields) as long as a HT frame can be defined (i.e., for subluminal shocks). In the results below we have chosen small magnetic field strengths so that the shock compression ratio does not depend on obliquity in order to minimize the confusion with injection efficiencies.

### 4.1. Acceleration Time

Figure 1 shows direct comparisons between the analytic results of Eq. (4) (smooth curves) and the Monte Carlo simulation (histograms) as a function of particle energy for a high Mach number shock of obliquity $\Theta_{\mathrm{Bn1}} = 60°$ and two values of the parameter $\eta = \lambda/r_g$. This clearly indicates, as do other simulation runs not depicted, that the simulation gives essentially identical results to Eq. (4) for all $\eta$, as long as $v/u_{1_{\mathrm{HT}}} \gg 1$, i.e. when the diffusion approximation that leads to Eq. (1) is valid. It should be noted that as $\Theta_{\mathrm{Bn1}}$ approaches $90°$, the condition for the validity of Eq. (1) becomes more severe (i.e. it is pushed to higher energies) and a limit is reached where $u_{1_{\mathrm{HT}}} > c$. We have not extended our results to these superluminal cases and refer readers to Jokipii and Terasawa (1995) for results on quasi-perpendicular shocks. The differences in $\tau_a$ at low energies are in the regime where the diffusion approximation and therefore Eq. (1) are invalid. Therefore these differences are unimportant if only the maximum acceleration time is considered, but can be significant if ionization or other processes which produce energy losses at low energies are important. In these cases, the Monte Carlo simulation can provide an effective way to study injection.

The Monte Carlo data points for shocks of obliquities $\Theta_{\mathrm{Bn1}} = 20°$ and $\Theta_{\mathrm{Bn1}} = 60°$ are shown in Fig. 3, which depicts the acceleration time $\tau_a$ divided by the final energy of the particle. The Monte Carlo data points are obtained by least squares fitting of histograms like those depicted in Fig. 1 at the highest energies. The comparisons illustrated in Fig. 3 are typical of those obtained for a full range of shock parameters and verify that the Monte Carlo simulation is in excellent agreement with Eq. (4) for the entire $\eta - \Theta_{\mathrm{Bn1}}$ phase space (the statistical uncertainties in the Monte Carlo results are smaller than the dots). While this comparison is made for high Mach number shocks, it is true of low $M_1$ also. Note that even near the Bohm limit, where the scattering length is of the order of a gyroradius and consecutive scatterings become somewhat correlated in phase, this comparison indicates that Eq. (3) is clearly applicable to our simulation; this is not surprising since the basic assumption of the kinetic theory approach is simply that particle momenta are isotropized upon scatterings.

### 4.2. Injection efficiency

We now investigate the injection efficiency as a function of Mach number, $\Theta_{\rm Bn1}$, and degree of cross-field scattering (i.e., $\eta$). The first problem is to define injection efficiency considering that no consensus exists concerning this. Since we are doing test-particle calculations, all spectra will become power-laws, with a spectral index which depends solely on the shock compression ratio, above some energy such that $v \gg u_{1\rm HT}$. Since $u_{1\rm HT}$ increases with $\Theta_{\rm Bn1}$ for a given $u_1$, the energy where the power-law begins will also increase with $\Theta_{\rm Bn1}$. To obtain a consistent measure of the injection efficiency for a shock with some $\Theta_{\rm Bn1}$ compared to the same shock at $0°$, we define the ratio of the *integral* number density of particles $n_{\Theta_{\rm Bn1}}(E > E_{\rm eff})$ accelerated to an energy $E_{\rm eff}$ or beyond (above the energy where the power-law is obtained) to the integral number density $n_0(E > E_{\rm eff})$ accelerated to the same energy or beyond in a $\Theta_{\rm Bn1} = 0°$ shock. This measure, which is therefore defined as

$$\epsilon(\Theta_{\rm Bn1}, E > E_{\rm eff}) \equiv \frac{n_{\Theta_{\rm Bn1}}(E > E_{\rm eff})}{n_0(E > E_{\rm eff})} \quad , \tag{7}$$

will be virtually independent of $E_{\rm eff}$ and gives a consistent measure of the *relative* injection efficiency for shocks with a *given compression ratio*. As mentioned above, even though our simulation works for all Mach numbers, we perform the following comparisons for cases where the magnetic field pressure is insignificant so that the compression ratio is virtually independent of the obliquity $\Theta_{\rm Bn1}$. This limit provides informative comparisons of efficiencies for shocks of different obliquities without the largely unenlightening encumbrance of magnetic field dependences. Absolute efficiencies, defined via the integral distribution as a ratio of non-thermal to thermal particles, are used by Baring, Ellison and Jones (1993). It must be noted, also, that since these are test-particle calculations, care must be taken when using $\epsilon$ as a measure of efficiency: true relative efficiencies can only be obtained with self-consistent calculations which include the backreaction of accelerated particles on the shock structure.

In Fig. 5 we show a set of integral density spectra for $M_1 = 100$, $\eta = 100$, and a range of shock obliquities. Our efficiency is calculated for $E_{\rm eff} = 1000$ keV, and is determined by the relative normalizations at that energy; since the spectral indices at these energies for the different obliquities are approximately equal, this determination is equivalent to Eq. (7). Again, here $B_1 = 10^{-8}$ G is small enough so that $r$ does not depend on $\Theta_{\rm Bn1}$. Fig. 5 clearly shows how the injection efficiency drops as $\Theta_{\rm Bn1}$ increases and, for this choice of parameters, we find that injection cuts off dramatically above $\Theta_{\rm cut} \simeq 35°$, a case that has poorer statistics. This cutoff effect, which occurs when scattering is weak (i.e., $\eta$ large), was explained in Baring, Ellison and Jones (1994). For cold upstream plasmas, particles convect along the field lines into the shock at speed $u_{1\rm HT}$ in the HT frame, and upon scattering in the downstream region (there is no reflection at the shock) assume a fluid frame speed of $v_{\rm F} = u_{1\rm HT} - u_{2\rm HT}$. If this speed is less than the downstream flow speed $u_{2\rm HT}$, the particles can never return to the shock and the acceleration process is shut off. Remembering that $u_{1x} = u_{1\rm HT}\cos\Theta_{\rm Bn1}$, $u_{2x} = u_{2\rm HT}\cos\Theta_{\rm Bn2}$ and that the compression ratio is given by $r = u_{1x}/u_{2x}$, the condition for this termination of injection of particles from the upstream gas into the diffusive acceleration process is $\cos\Theta_{Bn1}/\cos\Theta_{Bn2} \geq r/2$; for strong shocks ($r \approx 4$), injection then truly ceases for obliquities $\Theta_{Bn1} \gtrsim 30°$. Note also that higher obliquities imply higher $u_{1_{HT}}$, which means that the diffusion approximation is valid and the power-law distribution is achieved at higher energies than for plane-parallel shocks.

Lowering the Mach number increases the obliquity at which this cutoff occurs; strengthening the scattering toward the Bohm limit eventually eliminates this cutoff of acceleration, as is evident in Fig. 6, where we have plotted $\epsilon(\Theta_{Bn1}, E > E_{eff})$ versus $\Theta_{Bn1}$ for a number of cases, varying $M_1$ and $\eta$. In all cases, $B_1$ is small enough so that $M_1 \simeq M_{s1}$ and the compression ratio is independent of obliquity. Clearly, the relative injection efficiency as a function of $\Theta_{Bn1}$ is a strongly decreasing function of both $M_1$ and $\eta$. At low $M_1$, the upstream temperature is higher (we use a fixed value of $u_1 = 500$ km s$^{-1}$ in all cases) than in cold upstream plasma cases, and higher energy particles are present. These are more readily reflected (and energized) on first encountering the shock, or can end up downstream with more energy and a correspondingly greater chance of returning to the shock to be accelerated further. Hence low Mach number shocks are more efficient accelerators in this test-particle picture. When $\eta$ is low, cross-field diffusion is stronger, thereby increasing the flux of particles that can return to the shock from downstream, even for large obliquities. Hence strong scattering enhances the acceleration efficiency and removes the cutoff that appears when $\eta \gg 1$. Note that such strong scattering is necessary (Baring et al. 1995) in order for spectral data from the Monte Carlo technique to match Ulysses observations of protons at interplanetary shocks in the inner heliosphere.

When Figs. 2 and 6 are compared, we see that for a given set of parameters, the relatively sharp cutoff of $\epsilon$ at some $\Theta_{Bn1}$ translates to a maximum enhancement in acceleration rate over the $\Theta_{Bn1} = 0$ case. For example, shocks with $M_1 \gtrsim 100$ and with $\eta \gtrsim 100$ have a cutoff at $\Theta_{Bn1} \simeq 30°$ implying a maximum acceleration rate enhancement of a factor of about 4. Shocks with $M_1 \sim 3$ and $\eta \gtrsim 10$ have a cutoff at $\Theta_{Bn1} \simeq 70°$ that gives an optimal acceleration enhancement of about 40. Even with $\eta = 3$, the maximum acceleration rate enhancement for $M_1 = 3$ is only a factor of about 20. Clearly there is a trade-off between acceleration that is fast and acceleration that is efficient, a feature that we emphasize here because of its importance to astrophysical models that use diffusive acceleration at oblique shocks. We believe this work is the first to quantify this trade-off, and our method is ideally suited to assess when the acceleration process is optimized.

## 5. CONCLUSIONS

In this paper we have investigated the particle acceleration rate in oblique shocks using our Monte Carlo technique, and confirm previous analytic results obtained using the diffusion approximation (i.e., Jokipii 1987; Ostrowski 1988b), namely that highly oblique shocks accelerate particles more rapidly than quasi-parallel ones when the scattering is weak. Following Jokipii (1987), we have assumed that the scattering mean free path is proportional to gyroradius (i.e., $\lambda = \eta r_g$) with the constant parameter $\eta$ determining the strength of cross-field scattering. This assumption implies that the acceleration time to a given energy (and for a given set of shock parameters) is proportional to $\eta$ so that shocks with strong scattering (i.e., small $\eta$) accelerate particles more rapidly than ones with weak scattering. If scattering is strong ($\eta \sim 1$), the variation in acceleration rate with shock obliquity is modest; only when $\eta \gtrsim 10$ are gains larger than about 100 obtained and then only for a very restricted range in $\Theta_{Bn1}$ near 90° (see Figs. 2 or 3).

Being unencumbered by the restrictions of the diffusion approximation, our simulation approach also produces spectral information on the population and efficiency of accelerated particles. In this paper we have been able to quantify the relationship between scattering efficiency, Mach number, and obliquity. Specifically, we find that in addition to affecting the acceleration rate, the amount of cross-field diffusion strongly influences the injection efficiency in high Mach number shocks. If $\eta$ is large ($\gtrsim 100$), shocks with Mach numbers greater than about 100 (typical of young SNRs) do not inject significant numbers of thermal particles unless $\Theta_{Bn1} \lesssim 40°$, and even if the Mach number is as low as 3, injection occurs only if $\Theta_{Bn1} \lesssim 60°$ (see Fig. 6). Such low values of $\Theta_{Bn1}$ restrict any enhancement in the acceleration rate to less than about 20 times that of parallel shocks. For strong scattering (small $\eta$) the situation is somewhat different, with strong injection occurring even at high obliquity. For $\eta = 3$, the injection efficiency is $\gtrsim 10\%$ of the parallel shock value for $\Theta_{Bn1} \lesssim 80°$. However, for this $\eta$, the increase in acceleration rate over the $\Theta_{Bn1} = 0°$ rate is still less than about a factor of 20. In general, the results presented here indicate a trade-off between acceleration rapidity and efficiency, which is most extreme in shocks of high Mach number where large increases in acceleration rates are probably not effectively realized.

Our Monte Carlo technique clearly avoids some of the limitations of pure analytic methods, which generally cannot treat injection from thermal energies (though progress has been made in this direction: see Kang and Jones 1995), and also those of plasma simulations, which can model injection, but are severely limited by computational requirements and may not accurately describe cross-field diffusion in three dimensions. Even though we here model an infinite plane shock, our scattering technique is intrinsically three-dimensional and computationally efficient. We caution that the results presented here are for a particular type of scattering (i.e., large-angle), and are a test-particle calculation that does not include the effects of the backpressure of accelerated particles on the shock structure. We believe that the use of the large-angle scattering approximation is appropriate to a variety of astrophysical situations, for the reasons cited above. Clearly, pitch-angle diffusion and large-angle scattering will almost certainly give different injection efficiencies, at least in highly oblique shocks. We know that the acceleration process in relativistic shocks is very much dependent on the type of scattering assumed, where all particles have speeds not much greater than the flow speed and the spectral index of the power-law depends on the scattering mode (e.g., Ellison, Jones, and Reynolds 1990). However, we do not expect the effect of anisotropy to be as great in non-relativistic flows (unless $\Theta_{Bn}$ is such that the shock is nearly superluminal) since the majority of particles which scatter and cross the shock more than once to become superthermal have speeds considerably greater than the flow speed. These particles, while not necessarily isotropic, will be more isotropic than particles in relativistic flows. We are currently generalizing the code to include pitch-angle diffusion and will quantify the effect the scattering mode has on injection in a future paper.

The nonlinear effects from the backpressure of accelerated particles will also influence injection rates, and may cause the shock acceleration efficiency to self-regulate (as seen in parallel shocks, e.g., Eichler 1984; Ellison and Eichler 1984) so that the shock puts a large fraction of the total energy into energetic particles even if the number of particles injected from the background is quite small. Such nonlinear modifications are needed in applications to strong shocks such as the earth's bow shock (e.g. Ellison, Möbius, and Paschmann 1990), however we note that the test-particle simulations performed here should be adequate for the low Mach number, weak interplanetary shocks studied by Baring et al. (1995).

It must also be noted that we have only considered the case where injection all comes from a thermal background. If, in addition, energetic seed particles are present, such as pick-up ions in interplanetary shocks, they can be further accelerated by highly oblique shocks without the restrictions imposed by injection efficiency described here. It is also possible that in most actual shocks, such as SNRs, there will be regions of varying obliquity and the quasi-parallel portions may inject and energize thermal particles which then propagate into oblique regions of the shock where rapid acceleration can take place. In environments where this effect is important, complete models including realistic geometry will be necessary before even order of magnitude estimates of the acceleration efficiency can be made.

We thank Prof. Stephen Reynolds for many helpful discussions, and Prof. Randy Jokipii for reading the manuscript. MGB thanks the National Research Council for its sponsorship during the period in which the work for this paper was completed. This research was supported under the NASA Space Physics Theory Program.

# REFERENCES

Axford, W. I.: 1965 *Planet. Space Sci.* **13**, 115

Baring, M. G., Ellison, D. C. and Jones, F. C.: 1993 ApJ **409**, 327

Baring, M. G., Ellison, D. C. and Jones, F. C.: 1994 ApJS **90**, 547

Baring, M. G., Ellison, D. C. and Jones, F. C.: 1995 *Adv. Space Sci.* **8/9**, 397

Baring, M. G., Ogilvie, K. W., Ellison, D. C. and Forsyth, R.: 1995 *Adv. Space Sci.* **8/9**, 388

Baring, M. G. and Riitano, A. 1995 in preparation.

Begelman, M. C. and Kirk, J. G.: 1990 ApJ **353**, 66

Bell, A. R.: 1978 MNRAS **182**, 147

Blandford, R. D. and Eichler, D.: 1987 *Phys. Rep.* **154**, 1

Blandford, R. D. and Ostriker, J. P.: 1978 ApJ **221**, L29

Burgess, D.: 1989 *Geophys. Res. Lett.* **16**, 163

Decker, R. B.: 1988 *Space Sci. Rev.* **48**, 195

Decker, R. B. and Vlahos, L.: 1985 *J. Geophys. Res.* **90**, 47

de Hoffmann, F. and Teller, E.: 1950 *Phys. Rev.* **80**, 692

Drury, L. O'C: 1983 *Rep. Prog. Phys.* **46**, 973

Eichler, D.: 1984 ApJ **277**, 429

Eichler, D.: 1985 ApJ **294**, 40

Ellison, D. C.: *J. Geophys. Res.* **85**, 9029

Ellison, D. C. and Eichler, D.: 1984 ApJ **286**, 691

Ellison, D. C., Jones, F. C., and Reynolds, S. P.: 1990 ApJ **360**, 702

Ellison, D. C., Möbius, E., and Paschmann, G.: 1990 ApJ **352**, 376

Forman, M. A., and Morfill, G.: 1979, in *Proc. 16th Int. Cosmic Ray Conf.* (Kyoto), **5**, 328.

Forman, M. A., Jokipii, J. R. and Owens, A. J.: 1974 ApJ **192**, 535

Giacalone, J., Burgess, D. and Schwartz, S. J.: 1992 in Proc. 26th ESLAB Symposium: *Study of the Solar-Terrestrial System*, (ESA Special Publication, Noordwijk) p. 65

Giacalone, J., Burgess, D., Schwartz, S. J. and Ellison, D. C.: 1993 ApJ **402**, 550

Giacalone, J. and Jokipii, J. R.: 1994 ApJ **430**, L137

Hoppe, M. M., Russell, C. T., Frank, L. A., Eastman, T. E. and Greenstadt, E. W.: 1981 *J. Geophys. Res.* **86**, 4471

Jokipii, J. R.: 1987 ApJ **313**, 842

Jokipii, J. R., Kóta J. and Giacalone, J.: 1993 *Geophys. Res. Lett.* **20**, 1759

Jokipii, J. R., and Morfill, G. E.: 1985 ApJ **290**, L1

Jokipii, J. R., and Morfill, G.: 1987 ApJ **312**, 170

Jokipii, J. R., and Terasawa, T.: 1995 ApJ to be submitted.

Jones, F. C. and Ellison, D. C.: 1991 *Space Sci. Rev.* **58**, 259

Kang, H. and Jones, T. W.: 1995 ApJ in press.

Kucharek, H., Fujimoto, M. and Scholer, M.: 1993 *Geophys. Res. Lett.* **20**, 173

Lagage, P. O., and Cesarsky, C. J.: 1983 A&A **125**, 249

Max, C. E., Zachary, A. L. and Arons, J.: 1988, in Proc. Varenna School and Workshop *Plasma Astrophysics* , ESA SP-285, p. 45.

Ostrowski, M.: 1988a A&A **206**, 169

Ostrowski, M.: 1988b MNRAS **233**, 257

Ostrowski, M.: 1991 MNRAS **249**, 551

Quest, K. B.: 1988 *J. Geophys. Res.* **93**, 9649

Scholer, M., Fujimoto, M. and Kucharek, H.: 1993 *J. Geophys. Res.* **98**, 18,971

Takahara F. and Terasawa, T.: 1991 in Proc. ICRR International Symposium: *Astrophysical Aspects of the Most Energetic Cosmic Rays*, eds. Nagano, M., Takahara, F. (World Scientific, Singapore) p. 291

Terasawa, T.: 1979 *Planet. Space Sci.* **27**, 193

Winske, D. and Quest, K. B.: 1988 *J. Geophys. Res.* **93**, 9681

Winske, D., Omidi, N., Quest, K. B. and Thomas, V. A.: 1990 *J. Geophys. Res.* **95**, 18,821

Zachary, A. L.: 1987, Ph.D Thesis, Lawrence Livermore National Laboratory, University of California.

---

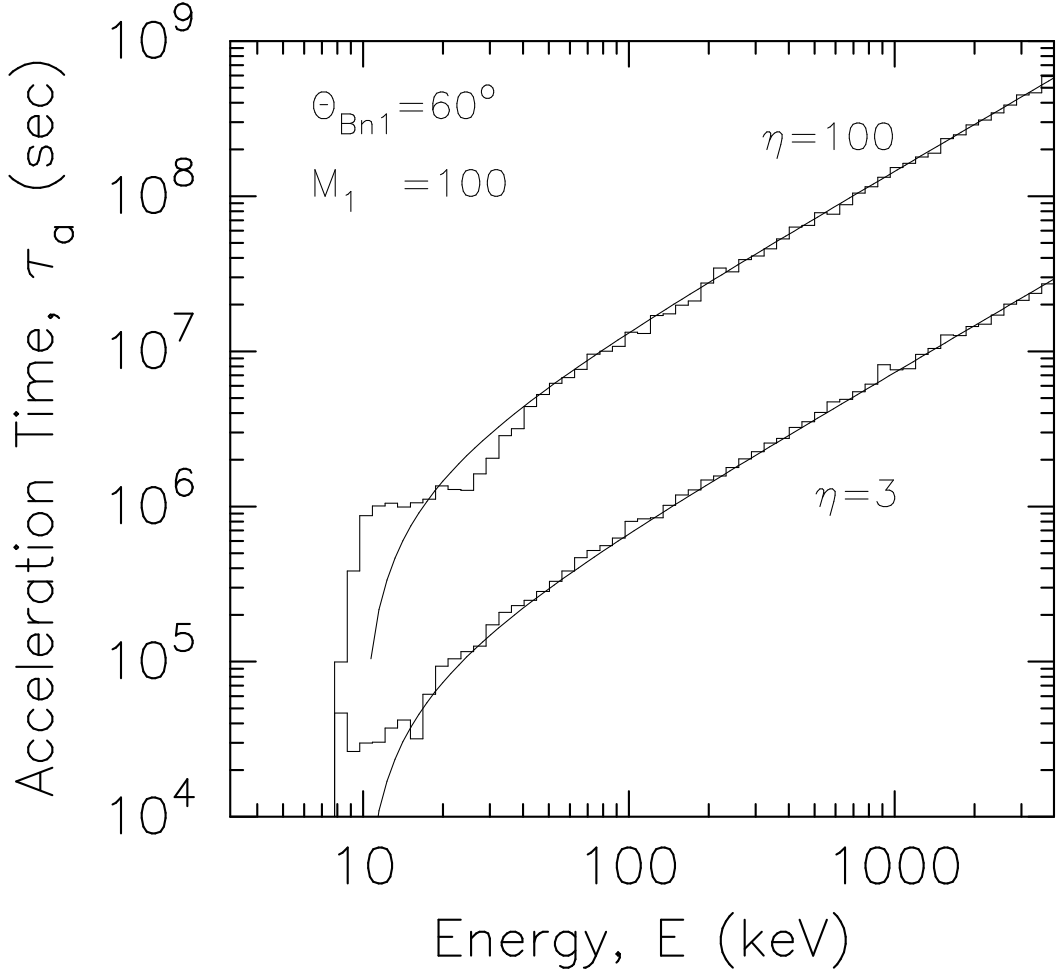

Fig. 1.— The average time $\tau_a$ (in seconds) taken to accelerate particles to an energy $E$, for $\Theta_{Bn1} = 60°$ and for $\eta = 3$ and $100$. The smooth curves are obtained from the analytic result in Eq. (4), and the histograms are obtained from the Monte Carlo simulation; they agree at the highest energies, where the diffusion approximation that is assumed in the derivation of Eq. (4) is valid. In all cases, $u_1 = 500$ km s$^{-1}$, upstream temperature $T_1 = 1.8 \times 10^3$ K, $B_1 = 10^{-8}$ G, and $E_i = 10$ keV. These input parameters yield an upstream Mach number, $M_1 \simeq M_{s1} = 100$ and a compression ratio, $r \simeq 4$.

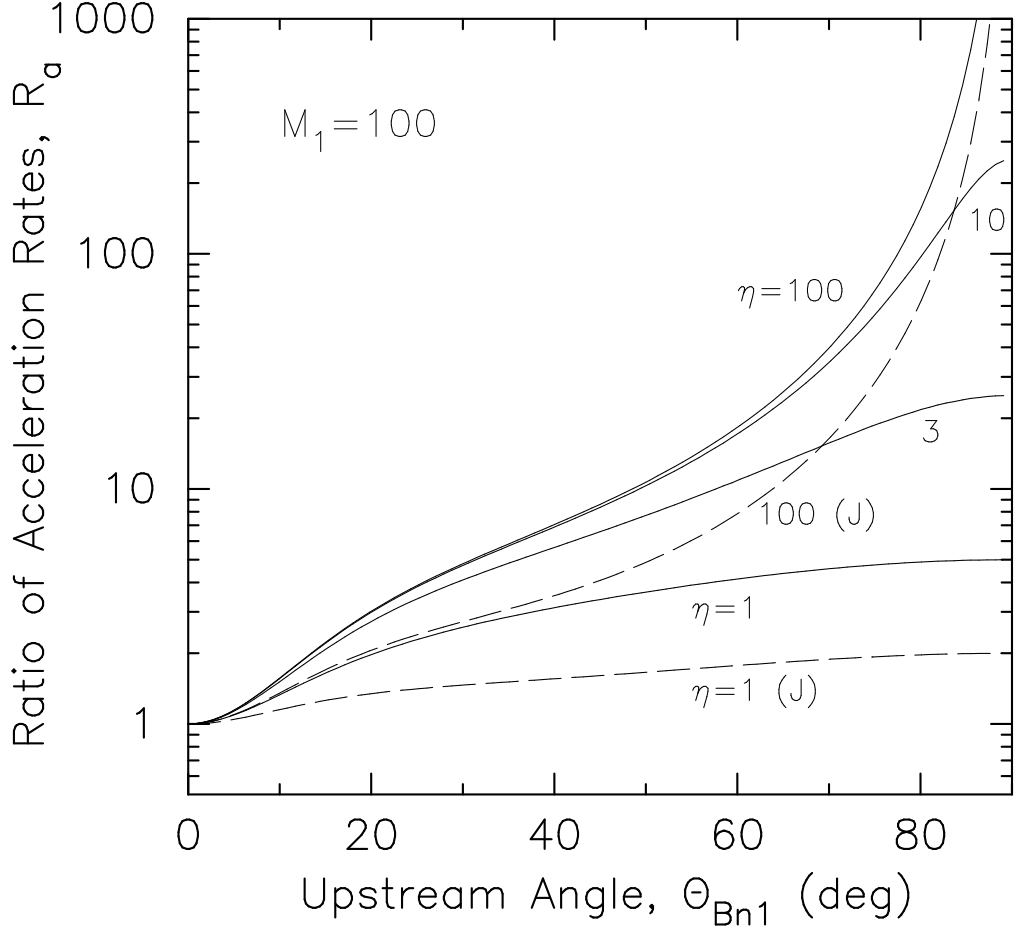

Fig. 2.— The ratio of acceleration rates $R_a$, comparing the case of parallel shocks to shocks with $\Theta_{Bn1}$ specified along the abscissa [see the form in Eq. (5)]. The solid lines represent the ratio when determined using the formula in Eq. (4), with $\eta = 1, 3, 10$, and 100, as labelled. The dashed lines are from Jokipii's (1987) prescription for $\tau_a$ with $\eta = 1$ and 100 (labelled with a J), and fall below our results for the reason given in the text. In all cases, the values of $u_1$, $T_1$, and $B_1$ are those used for Fig. 1, and again $r = 4$ and $M_1 = 100$.

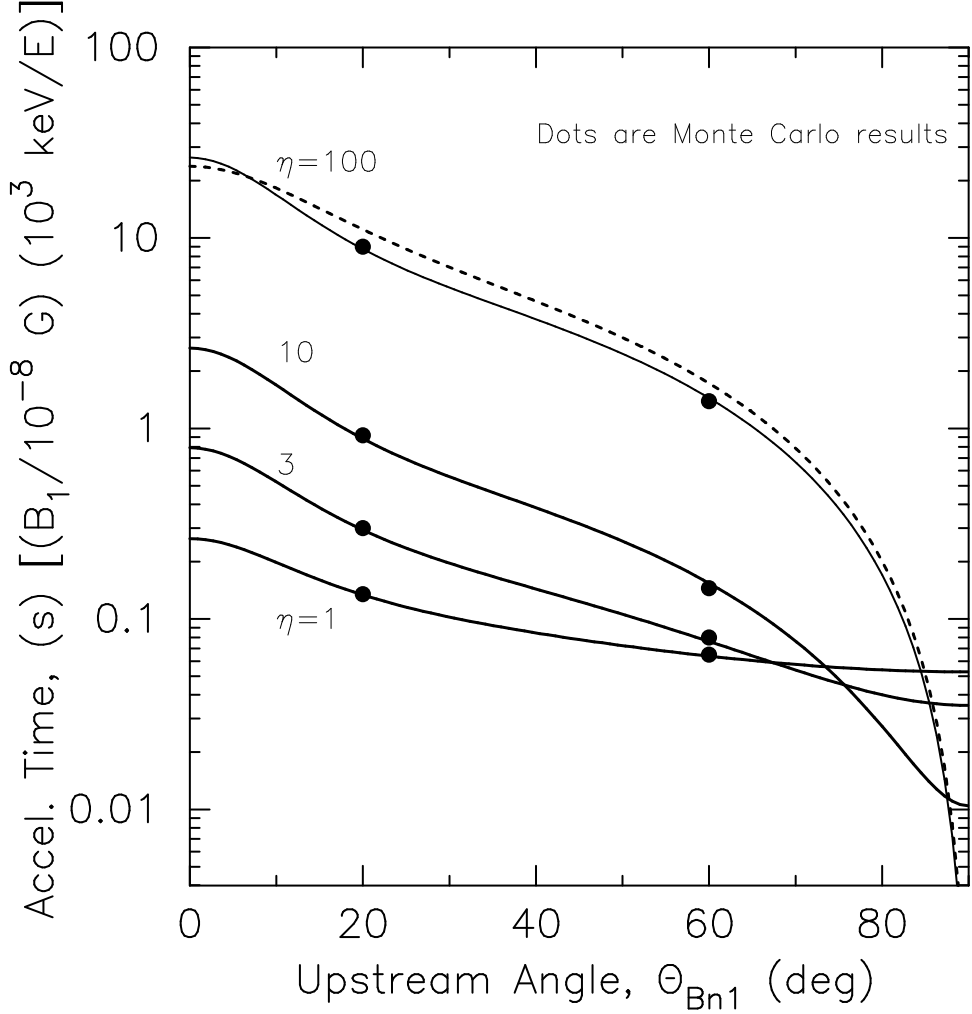

Fig. 3.— The acceleration time, as defined in Eq. (4), as a function of shock obliquity for various $\eta$ as labelled, for particles accelerated to energy $E = 10^3$ keV. The solid curves are $M_1 = 100$ cases with shock parameters as in Figs. 1 and 2. The dashed curve is for $M_1 = 3$ and $\eta = 100$, indicating how insensitive the time is to Mach number when $\eta$ is large. The data points at $\Theta_{Bn1} = 20°$ and $\Theta_{Bn1} = 60°$ are obtained from the Monte Carlo simulation for $M_1 = 100$ shocks: they are obtained by least squares fitting of histograms like those depicted in Fig. 1, and indicate very close agreement for all $\eta$, even near unity. Results are normalized by the upstream magnetic field strength $B_1$ and the maximum energy gained by the particle.

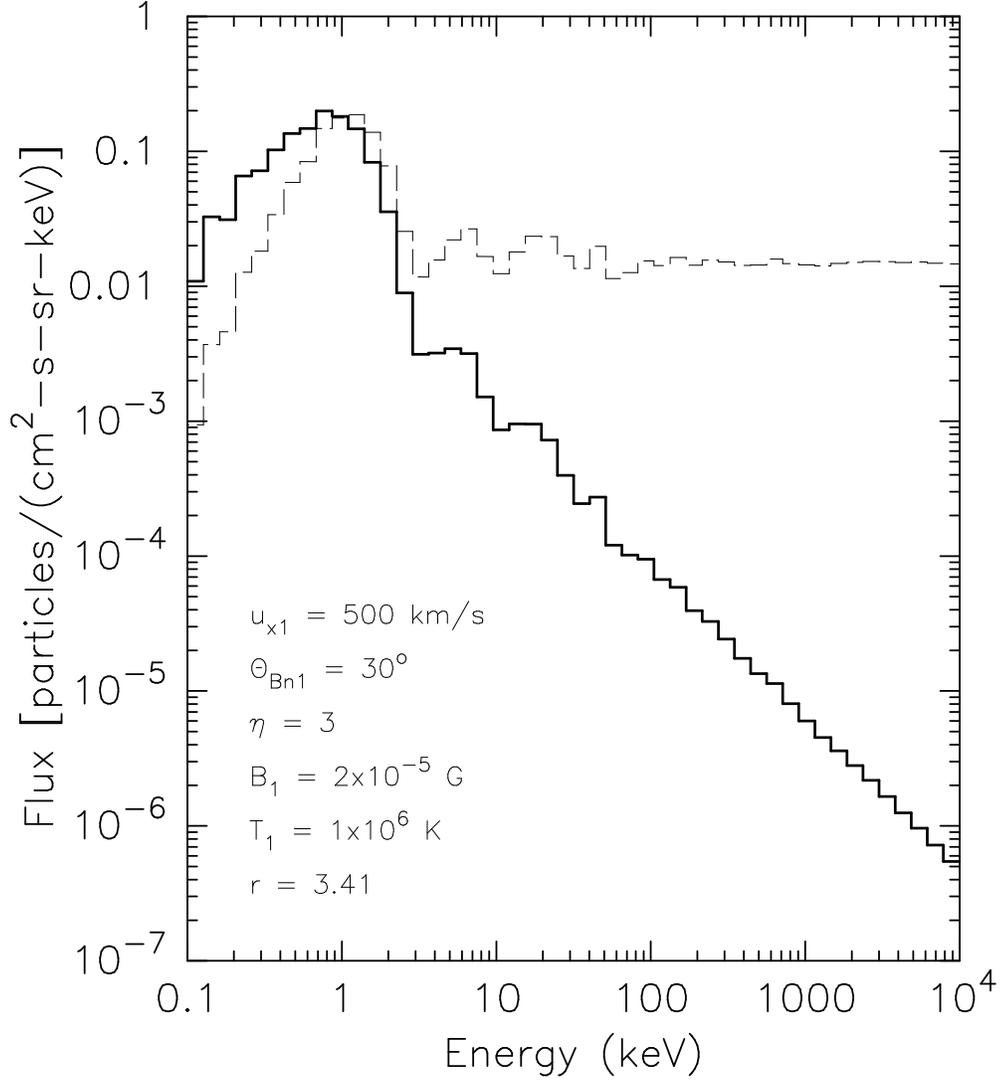

Fig. 4.— Omni-directional, differential flux versus energy for a shock with a dynamically important magnetic field strength. Depicted are both the Monte Carlo spectrum $dJ/dE$ (solid histogram, calculated in the shock frame) and also $E^\sigma \, dJ/dE$ (dashed histogram), where $\sigma = [(r+2)/(r-1)]/2$ is the canonical theoretical spectral index (e.g. Drury 1983, Jones and Ellison, 1991). Clearly, at energies where $v \gg u_{1\mathrm{HT}}$, the simulated spectrum agrees impressively with the theory. The flux is normalized to one incoming particle per cm$^2$ per second.

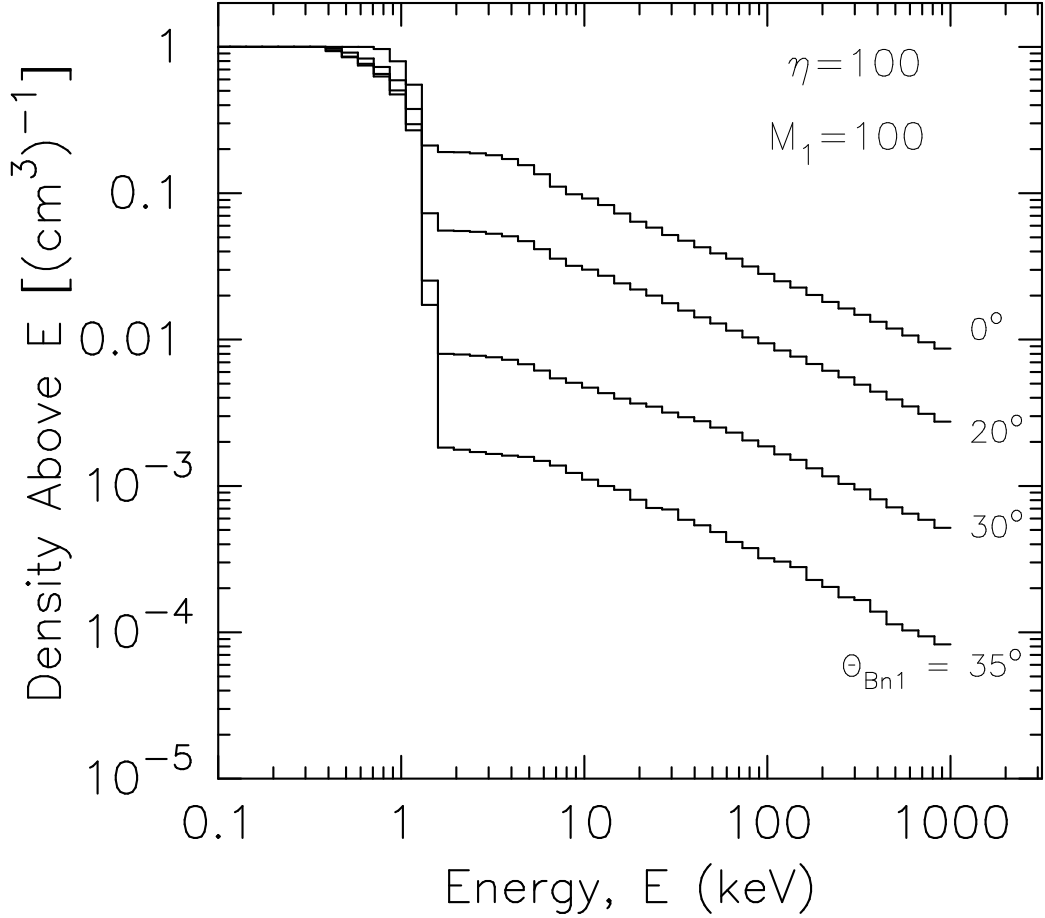

Fig. 5.— Integral density distributions for particles accelerated in a shock of upstream Mach number $M_1 = 100$ and scattering parameter $\eta \equiv \lambda/r_g = 100$ (weak cross-field diffusion), for different obliquities $\Theta_{Bn1}$ as labelled. The field strength was chosen low ($B_1 = 10^{-8}$ G) to maintain a constant compression ratio ($r \approx 4$) for all obliquities. The comparative efficiencies of injection can be represented by the relative normalizations of the power-law portions of the integral distributions. The distributions are calculated in the shock frame beyond the downstream boundary to the simulation. Apart from statistical noise, all cases obtain the same power law slope at high energies. Note that differential distributions for this shock regime, but with a guiding center approach in the simulation, are depicted in Baring, Ellison and Jones (1993).

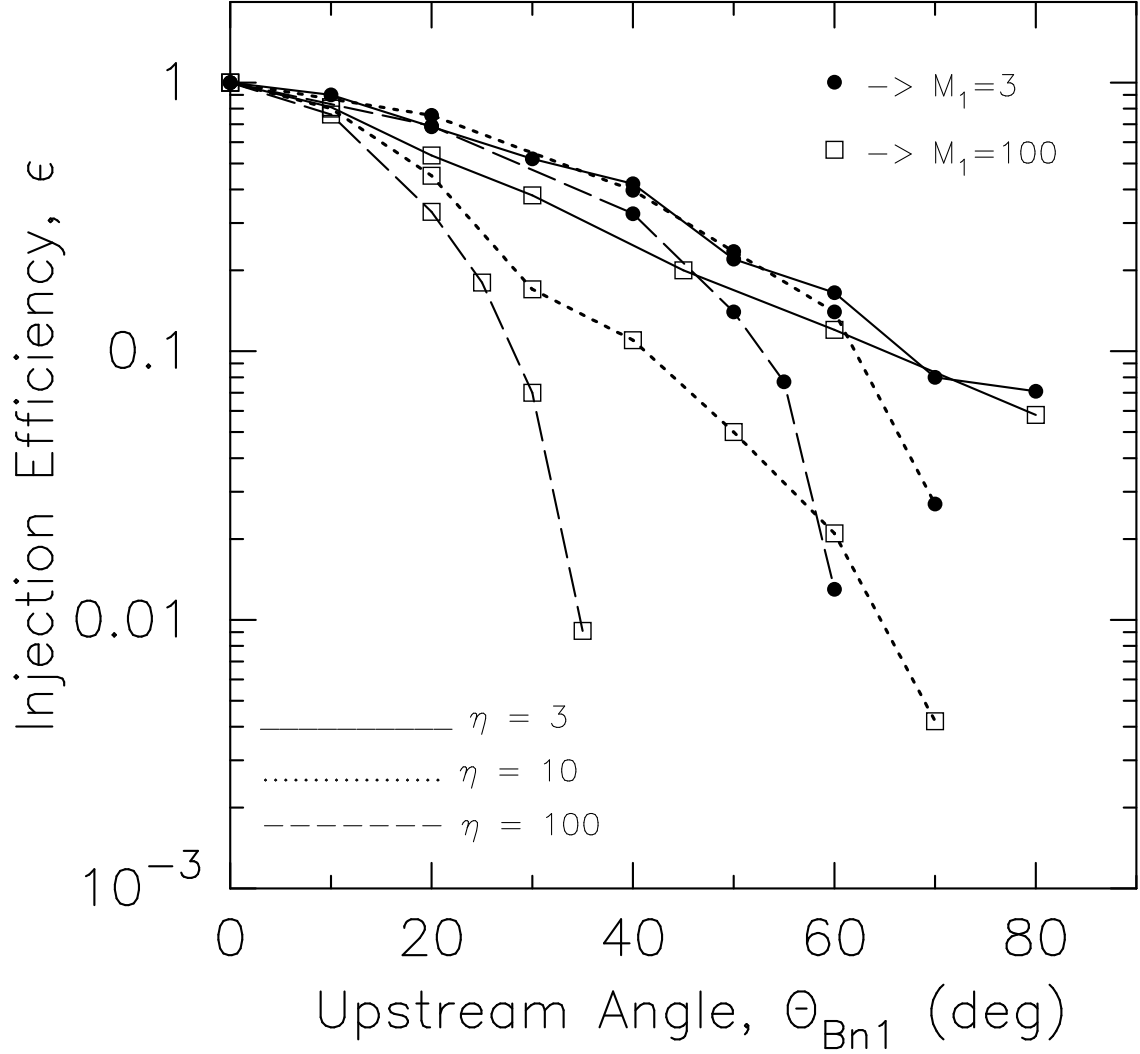

Fig. 6.— The injection efficiency, defined in Eq. (7), as a function of shock obliquity for different $\eta$, as labelled, and Mach numbers $M_1 = 100$ (squares) and $M_1 = 3$ (dots). In general, the weaker the scattering (large $\eta$) and the larger the Mach number, the more rapidly the efficiency falls off with $\Theta_{Bn1}$. For the $\eta = 100$, $M_1 = 100$ case, essentially no injection occurs for $\Theta_{Bn1} > 35°$. The jitter in the various "curves" defines the statistical uncertainty associated with efficiency determinations from spectra like those in Fig. 5.